\documentstyle[romp36]{article}

\def\q1{{q^{-1}}}
\def\qq1{{q-q^{-1}}}

\def\nq{{n_{i}}}
\newcommand{\bfm}[1]{\mbox{\boldmath${#1}$}}
\begin{document}

\title{q-deformed structures and generalized thermodynamics}
\author{A. Lavagno$^{1,2}$, A.M. Scarfone$^{1,3}$ and P.Narayana Swamy $^4$\\
{\small ~~~~~~~~~~}\\
{\small $^1$Dipartimento di Fisica, Politecnico di Torino, Italy}\\
{\small $^2$Istituto Nazionale di Fisica Nucleare, Sezione di Torino, Italy}\\
{\small $^3$Istituto Nazionale di Fisica della Materia, Sezione di Torino, Italy}\\
{\small $^4$Southern Illinois University, Edwardsville, IL 62026, USA}
} \maketitle

\begin {abstract}

On the basis of the recently proposed formalism [A. Lavagno and
P.N. Swamy, Phys. Rev. E {\bf65}, 036101 (2002)], we show that the
realization of the thermostatistics of $q$-deformed algebra can be
built on the formalism of $q$-calculus. It is found that the
entire structure of thermodynamics is preserved if we use an
appropriate Jackson derivative instead of the ordinary
thermodynamic derivative. Furthermore, in analogy with the quantum
$q$-oscillator algebra, we also investigate a possible
$q$-deformation of the classical Poisson bracket in order to
extend a generalized $q$-deformed dynamics in the classical
regime.

PACS number(s): 03.65.-w, 05.30.-d, 45.20.-d, 02.20.Uw 
\end {abstract}


\section{Introduction}

In the last few years quantum algebra and quantum groups have been
the subject of  intensive research in several physical fields.
Although some properties of these structures still deserve more
study in order to provide full clarity, it was quite clear from
the beginning that the basic features of these generalized
theories are strongly connected with a deep physical meaning and
not merely a mathematical exercise. Quantum algebra and quantum
groups, which first emerged in connection with the quantum inverse
scattering theory and statistical mechanics model \cite{Baxter},
where the quantum Yang-Baxter equation plays a crucial role, have
found recent important applications in many physical and
mathematical problems, such as conformal field theory, integrable
systems, non-commutative geometry, knot theory, thermostatistical
models and in a wide range of applications envisaged, from cosmic
strings and black holes to the fractional quantum Hall effect and
high-$T_c$ superconductors \cite{wil,lerda,alva}.

From the seminal work of Biedenharn \cite{bie} and Macfarlane
\cite{mac} it was clear that the $q$-calculus, originally
introduced at the beginning of this century by Jackson \cite{jack}
in the study of the basic hypergeometric function, it plays a
central role in the representation of the quantum groups
\cite{exton}. In fact it has been shown that it is possible to
obtain a ``coordinate" realization of the Fock space of the
$q$-oscillators by using the deformed Jackson derivative (JD) or
the so-called $q$-derivative operator \cite{flo,cele1,fink}. In
particular, in the paper of Celeghini {et al.} \cite{cele1}, the
quantum Weyl-Heisemberg algebra is studied in the frame of the
Fock-Bargmann representation and is incorporated into the theory
of entire analytic functions allowing a deeper mathematical
understanding of squeezed states, relation between coherent
states, lattice quantum mechanics and Bloch functions.

Furthermore, in the recent past there has been increasing emphasis
in quantum statistics different from that of standard bosons and
fermions. Since the pioneering work of Gentile and Green
\cite{genti,green}, there have been many extensions beyond the
standard statistics, among the others we may list the following:
parastatistics, fractional statistics, quon statistics, anyon
statistics and quantum groups statistics. In the literature there
are two principal methods of introducing an intermediate
statistical behavior. The first is to deform the quantum algebra
of the commutation-anticommutation relations thus deforming the
exchange factor between permuted particles. The second method is
based on modifying the number of ways of assigning particles to a
collection of states and thus the statistical weight of the
many-body system. One interesting realization of the first
approach is the study of exactly solvable statistical systems
which has led to the theory of the $q$-deformed harmonic
oscillator \cite{bie,mac}, based on the construction of SU$_q$(2)
algebra of $q$-deformed commutation or anticommutation relations
between creation and annihilation operators. Such an algebra opens
the possibility to study intermediate $q$-boson and $q$-fermion
statistical behavior \cite{chia,marmo,song} and to introduce a
generalized thermostatistics based on the formalism of the
$q$-calculus \cite{pre}. In this context, based on the formulation
of generalized statistical mechanics in connection with
$q$-deformed calculus, it has been pointed out that the thermal
average of an operator is closely related to quantum algebra
\cite{cele2} in a fully developed scheme.

The purpose of this paper is twofold: first, we want to show that
the realization of the thermostatistics of $q$-deformed algebra
can be built on the formalism of $q$-calculus and that the entire
structure of thermodynamics is preserved if we use the
(non-commutative) JD instead of the ordinary thermodynamic
derivative. Second, starting from the observation that the
creation and annihilation operators in the quantum $q$-deformed
SU$_q$(2) algebra correspond classically to non-commuting
coordinates in a $q$-phase-space and the commutation relation
between the standard quantum operators corresponds classically to
the Poisson bracket (PB), we want to introduce a $q$-deformation
of the PB ($q$-PB) in order to define a generalized $q$-deformed
dynamics in a non-commutative phase-space. The motivation for this
second goal lies in the fact that a full understanding of the
physical origin of $q$-deformation in classical physics is still
lacking because  it is not clear if there exists a classical
counterpart to the quantum groups. It is thus an outstanding
question:  how can classical dynamics be formulated such that,
upon canonical quantization, one recovers the deformed quantum
$q$-oscillator theory. The problem of a possible $q$-deformation
of  classical  mechanics was dealt with in Ref. \cite{Klimek}
where a $q$-PB has been obtained starting from a point of view
different  from the one adopted in this paper. As the quantum
$q$-deformation plays a crucial role in the interpretation of
several complex physical systems,  we expect that a classical
$q$-deformation of the dynamics can be very relevant in several
problems \cite{marmo1}. A remarkable example is the Tsallis
nonextensive thermostatistics \cite{tsallis}, based on a classical
deformation of the Boltzmann-Gibbs entropy, where the
thermodynamic functions, such as entropy and internal energy, are
deformed, but the whole structure of thermodynamics is preserved.
Many investigations are devoted to the relevance of such a
classical theory in several physical applications \cite{tsallis2}.

The main approach we shall follow to introduce the classical
correspondence of quantum $q$-oscillator is based on the following
idea. The (undeformed) quantum commutation relations are invariant
under the action of SU(2) and, as a consequence, the $q$-deformed
commutation relations are invariant under the action of SU$_q$(2).
Analogously, since the (undeformed) PB is invariant under the
action of the symplectic group Sp(1), we have to require that
$q$-PB must satisfy invariance over the action of the $q$-deformed
symplectic group Sp$_{q}$(1).

The paper is organized as follows. In Sec. 2 we review the
fundamental relations in the $q$-Heisenberg algebra of creation
and annihilation operators and we derive the statistical
distribution for $q$-boson gas. In Sec. 3 we introduce a
self-consistent prescription for the use of the JD in the
thermodynamical relations. In Sec. 4, we define the
non-commutative $q$-differential calculus in the $q$-deformed
phase-space invariant over the action of the GL$_q$(2) and in Sec.
5 we introduce the $q$-symplectic group and the $q$-PB. Finally,
we present our conclusions in Sec. 6.


\section{Realization of the $q$-boson algebra and thermal averages}

We shall briefly review the theory of  $q$-deformed bosons defined
by the $q$-Heisenberg algebra of creation and annihilation
operators of bosons introduced by Biedenharn and McFarlane
\cite{bie,mac}, derivable through a map from SU$_q$(2). The
$q$-boson algebra is determined by the following commutation
relations for annihilation and creation operators $a$, $a^{\dag}$
and the number operator $N$, thus (for simplicity we omit the
particle index)
\begin{equation}
[a,a]=[a^\dag,a^\dag]=0 \; , \ \ \ aa^\dag-q^2 a^\dag a =1 \; ,
\end{equation}
\begin{equation}
[N,a^\dag]= a^\dag \; , \ \ \ [N,a]=-a\; .
\end{equation}
The $q$-Fock space spanned by the orthornormalized eigenstates
$\vert n\rangle$ is constructed according to
\begin{equation}
\vert n\rangle=\frac{(a^\dag)^n}{\sqrt{[n]!}} \vert 0\rangle \; ,
\ \ \ a\vert 0\rangle=0 \; ,
\label{fock}
\end{equation}
where the $q$-basic factorial is defined as
\begin{equation}
[n]!=[n] [n-1] \cdots [1]
\label{brnf}
\end{equation}
and the $q$-basic number $[x]$ is defined in terms of the $q$-deformation parameter
\begin{equation}
[x]=\frac{q^{2x}-1}{q^2-1}\; .
\label{bn}
\end{equation}
In the limit $q \rightarrow 1$, the $q$-basic number $[x]$ reduces
to the ordinary number $x$ and all the above relations reduce to
the standard boson relations.

The actions of $a$, $a^\dag$ on the Fock state $\vert n \rangle$ are given by
\begin{eqnarray}
a^\dag \vert n\rangle &=& [n+1]^{1/2} \vert n+1\rangle\; , \\
a \vert n\rangle&=&[n]^{1/2} \vert n-1\rangle \; ,\\
N \vert n\rangle&=&n\vert n\rangle \; .
\end{eqnarray}
From the above  relations, it follows that $a^\dag a=[N]$,
$aa^\dag=[N+1]$.

We observe that the Fock space of the $q$-bosons has the same structure as
the standard bosons but with the replacement $n!\rightarrow [n]!$ .
 Moreover the number operator is not $a^\dag a$ but can be expressed as the
 nonlinear functional relation $N=f(a^\dag a)$ which can be
explicitly obtained formally in the closed form
\begin{equation}
N=\frac{1}{2 \log q} \log\Big (1+(q^2-1) a^\dag a \Big )\; .
\label{nop}
\end{equation}
The transformation from Fock observables to the configuration
space (Bargmann holomorphic representation) may be accomplished by
choosing \cite{flo,cele1,fink}
\begin{equation}
a^\dag\rightarrow x \; , \ \ \ a\rightarrow {\cal D}_x \; ,
\label{jd}
\end{equation}
where ${\cal D}$ is the JD \cite{jack} defined by
\begin{equation}
{\cal D}_x\,f(x)=\frac{f(q^2\,x)-f(x)}{(q^2-1)\,x}\; ,
\end{equation}
which reduces to the ordinary derivative  when $q$ goes to unity
and therefore, the JD occurs naturally in $q$-deformed structures.
In an analogous manner, we will see below that JD plays a crucial
role in the $q$-deformed classical mechanics also.

The thermal average of an operator is written in the standard form
\begin{equation}
\langle {\cal O}\rangle=\frac{ Tr \left ({\cal O} \, e^{-\beta H} \right )}{\cal Z}\; ,
\end{equation}
where $\cal Z$ is the grand canonical partition function defined as
\begin{equation}
{\cal Z}=Tr \left ( e^{-\beta H} \right )\; ,
\label{pf}
\end{equation}
and $\beta = 1/T$. Henceforward we shall set Boltzmann constant to unity.

By using the definition in Eq.(\ref{bn}) of the $q$-basic number, the mean value of
the occupation number $\nq$ can be calculated starting
from the relation
\begin{equation}\label{14}
[n_{i}]=\frac{1}{\cal Z}\, Tr\left ( e^{-\beta H} a^\dag_i
a_i\right ) \; .
\end{equation}
As outlined before, the consistency of this approach is warranted
by the fact that the thermal averages of an operator is related
with quantum algebra in a fully consistent scheme \cite{cele2}.

In the grand canonical ensemble, the Hamiltonian of the
non-interacting boson gas is expected to have the following form
$H=\sum_i (\epsilon_i-\mu) \, N_i$ where the index $i$ is the
state label, $\mu$ is the chemical potential and $\epsilon_i$ is
the kinetic energy in the state $i$ with the number operator
$N_i$. Therefore from Eq.(\ref{14}), after simple manipulations
the explicit expression for the mean occupation number can be
determined as
\begin{equation}
n_{i}=\frac{1}{2 \log q}
\log\left (\frac{z^{-1}e^{\beta\epsilon_i}-1}{ z^{-1}e^{\beta\epsilon_i}-q^2}
\right) \; ,
\label{nqi}
\end{equation}
where $z = e^{\beta \mu}$ is the fugacity. It is easy to see that the above equation
reduces to the standard Bose-Einstein distribution
when $q\rightarrow 1$. The total number of particles is given by $N=\sum_i\,n_i$.

\section{$q$-calculus in the deformed thermodynamic
 relations}

From the definition of the partition function, and the
Hamiltonian, it follows that the logarithm of the partition
function has the same structure as that of the standard boson
\begin{equation}
\log {\cal Z}=-\sum_i \log (1-z e^{-\beta\epsilon_i}) \; .
\end{equation}
This is due to the fact that we have chosen the Hamiltonian to be a linear function of the
number operator but it is  not linear in $a^\dag a$
as seen from Eq.(\ref{nop}). For this reason,
the standard  thermodynamic relations in the usual form are ruled out. It is verified,
for instance, that
\begin{equation}
N\ne z \, \frac{\partial}{\partial z} \log {\cal Z}\; .
\end{equation}
As the coordinate space representation of the $q$-boson algebra is
realized by the introduction of the JD (see Eq.(\ref{jd})), we
stress  that the key point of the $q$-deformed thermostatistics is
in the observation that the ordinary thermodynamic derivative with
respect to $z$, must be replaced by the JD \cite{pre}
\begin{equation}
\frac{\partial}{\partial z} \Longrightarrow {\cal D}_z \; .
\end{equation}
Consequently, the number of particles in the $q$-deformed theory can be derived from the relation
\begin{equation}
N=z \; {\cal D}_z \log {\cal Z}\equiv \sum_i n_i \; ,
\label{num}
\end{equation}
where $n_i$ is the mean occupation number expressed in Eq.(\ref{nqi}).

The usual Leibniz chain rule is ruled out for the JD and therefore derivatives encountered
in thermodynamics must be modified according to the
following prescription. First we observe that the JD applies only with respect to the variable
in the exponential form such as $z=e^{\beta \mu}$
or $y_i=e^{-\beta \epsilon_i}$. Therefore for the $q$-deformed case, any thermodynamic
derivative of functions which depend on $z$ or $y_i$ must
be converted to derivatives in one of these variables by using the ordinary chain rule and
then applying the JD with respect to the exponential
variable.  For example, the internal energy in the $q$-deformed case can be written as
\begin{equation}
U=-\left. \frac{\partial}{\partial\beta} \log {\cal Z} \right |_z=\sum_i
\frac{\partial y_i}{\partial\beta} \, {\cal D}_{y_i}\log(1-z\,y_i) \; .
\label{int}
\end{equation}
In this case we obtain the correct form of the internal energy
\begin{equation}
U=\sum_i \epsilon_i \, \nq\; .
\label{un}
\end{equation}
This prescription is a crucial point of our approach because this
allows us to maintain the whole structure of thermodynamics and
the validity of the Legendre transformations in a fully consistent
manner. For instance, in light of the above discussion, we have
the recipe to derive the entropy of the $q$-bosons which leads to
\begin{eqnarray}
S=-\left. \frac{\partial\Omega}{\partial T}\right |_\mu &&\equiv
\log {\cal Z} +\beta\sum_i\left.\frac{\partial\kappa_i}{\partial\beta}\right|_\mu
{\cal D}^{(q)}_{\kappa_i}\log(1-\kappa_i)\nonumber\\
&&=\log {\cal Z} +\beta U-\beta\mu N \; ,
\end{eqnarray}
where $\kappa_i=z\, e^{-\beta\epsilon_i}$,
$U$ and $N$ are the modified functions expressed in Eqs.(\ref{int}) and (\ref{num})
and $\Omega = - T \log {\cal Z}$ is the thermodynamic potential.
After some manipulations, we obtain the entropy as follows
\begin{equation}
S=\sum_i \Big \{ -\nq \, \log\, [\nq]+(\nq+1)\,  \log\, [\nq+1]- \log\, ([\nq+1]-[\nq])
\Big \} \; .
\label{entro}
\end{equation}

\section{Non-commutative differential calculus}

In the above sections we have seen that the non-commutative JD and
the $q$-calculus play a crucial role in the definition of the
$q$-deformed quantum mechanics and quantum thermodynamics. We
would now like to introduce an analogous $q$-deformation in
classical theory. Since the creation and annihilation operators in
the quantum $q$-deformed SU$_q$(2) algebra correspond classically
to non-commuting coordinates in a $q$-phase-space, in this section
we introduce the $q$-deformed plane which is generated by the
non-commutative elements $\hat x$ and $\hat p$ fulfilling the
relation \cite{Reshetikhin}
\begin{equation}
\hat p\,\hat x=q\,\hat x\,\hat p \ ,
\label{qplane}
\end{equation}
which is invariant under GL$_q$(2) transformations (see later).
Henceforward, for simplicity, we shall limit ourselves to consider
the two-dimensional case.

From Eq.(\ref{qplane}) the $q$-calculus on the $q$-plane can be
obtained formally through the introduction of the $q$-derivatives
$\hat\partial_x$ and $\hat\partial_p$ \cite{Wess}
\begin{eqnarray}
&&\hat\partial_p\,\hat p=\hat\partial_x\,\hat x=1 \ ,\\
&&\hat\partial_p\,\hat x=\hat\partial_x\,\hat p=0 \ .
\end{eqnarray}
They fulfill the $q$-Leibniz rule
\begin{eqnarray}
&&\hat\partial_p\,\hat p=1+q^2\,\hat
p\,\hat\partial_p+(q^2-1)\,\hat x\,\hat\partial_x \ ,\label{pp}\\
&&\hat\partial_p\,\hat x=q\,\hat x\,\hat\partial_p\ ,\label{px}\\
&&\hat\partial_x\,\hat p=q\,\hat p\,\hat\partial_x\ ,\label{xp}\\
&&\hat\partial_x\,\hat x=1+q^2\,\hat x\,\hat\partial_x \
,\label{xx}
\end{eqnarray}
together with the $q$-commutative derivative
\begin{equation}
\hat\partial_p\,\hat\partial_x=q^{-1}\,\hat\partial_x\,\hat\partial_p \ .
\end{equation}
It is easy to see that in the $q\rightarrow 1$ limit one recovers the ordinary commutative calculus.
Let us outline the asymmetric mixed derivative relations Eq.(\ref{pp}) and Eq.(\ref{xx}) in $\hat x$ and in $\hat p$.
These properties come directly from the non-commutative structure of the $q$-plane defined in Eq.(\ref{qplane}).

We recall now that the most general function on the $q$-plane can
be expressed as a polynomial in the $q$-variable $\hat x$ and
$\hat p$
\begin{equation}
f(\hat x,\,\hat p)=\sum_{i,j}c_{ij}\,\hat x^i\,\hat p^j \ .
\end{equation}
where we have assumed the $\hat x$-$\hat p$ ordering prescription
(it can be always accomplished by means of Eq. (\ref{qplane})).
Thus, taking into account Eqs.(\ref{pp})-(\ref{xx}), we obtain
the action of the $q$-derivatives on the monomials
\begin{eqnarray}
&&\hat\partial_x(\hat x^n\,\hat p^m)=[n]\,\hat x^{n-1}\,\hat p^m
\ ,\label{dxmon}\\
&&\hat\partial_p(\hat x^n\,\hat p^m)=[m]\,q^n\,\hat x^n\,\hat
p^{m-1} \ ,\label{dpmon}
\end{eqnarray}
where $[n]$ is the same $q$-basic number introduced in Eq.(\ref{bn}).

A realization of the above $q$-algebra and its $q$-calculus can be
accomplished by the replacements \cite{Ubriaco}
\begin{eqnarray}
&&\hat x\rightarrow x \ ,\label{x}\\
&&\hat p\rightarrow p\,D_x \ ,\label{p}\\
&&\hat\partial_x\rightarrow{\cal D}_x \ ,\label{dx}\\
&&\hat\partial_p\rightarrow{\cal D}_p\,D_x \ ,\label{dp}
\end{eqnarray}
where
\begin{equation}
D_x=q^{x\,\partial_x} \ ,\hspace{20mm}D_x f(x,\,p)=f(q\,x,\,p) \ ,
\end{equation}
is the dilatation operator along the $x$ direction (reducing to
the identity operator for $q\rightarrow 1$), whereas
\begin{equation}
{\cal D}_x=\frac{q^{2\,x\,\partial_x}-1}{(q^2-1)\,x} \
,\hspace{15mm}{\cal D}_p=\frac{q^{2\,p\,\partial_p}-1}{(q^2-1)\,p}\ ,
\end{equation}
are the JD with respect to $x$ and $p$. Their action on an
arbitrary function $f(x,\,p)$ is
\begin{equation}
{\cal D}_x\,f(x,\,p)=\frac{f(q^2\,x,\,p)-f(x,\,p)}{(q^2-1)\,x} \
,\hspace{15mm}{\cal
D}_p\,f(x,\,p)=\frac{f(x,\,q^2\,p)-f(x,\,p)}{(q^2-1)\,p} \ .
\end{equation}
Therefore, as a consequence of the non-commutative structure of
the $q$-plane, in this realization the $\hat x$ coordinate becomes
a $c$-number and its derivative is the JD whereas the $\hat p$
coordinate and its derivative are realized in terms of the
dilatation operator and JD.


\section{$q$-Poisson bracket and $q$-symplectic group}

After the formulation of the $q$-differential calculus, we are now
able to introduce a $q$-PB. As previously stated, since the
undeformed PB is invariant under the action of the undeformed
symplectic group Sp(1), we will assume, as a fundamental point,
that the $q$-PB must satisfy the invariance property under the
action of the $q$-deformed symplectic group Sp$_q$(1) with the
same  value of the deformed parameter $q$ used in the construction
of the quantum plane.

Let us start by recalling the classical definition of a 2-Poisson
manifold, which is a two dimensional Euclidean space $I\!\!R^2$
generated by the position and momentum variables $x\equiv x^1$ and
$p\equiv x^2$ and equipped with a PB. If $f(x,\,p)$ and $g(x,\,p)$
are smooth functions, the PB is defined as \cite{gold}
\begin{equation}
\Big\{f,\,g\Big\}=\partial_xf\,\partial_pg-
\partial_pf\partial_xg \
.\label{poisson}
\end{equation}
Eq. (\ref{poisson}) can be expressed in a compact form
\begin{equation}
\Big\{f,\,g\Big\}=\partial_{i}f\,J^{ij}\,\partial_{j}g \
,\label{poisson1}
\end{equation}
where $J$ is the unitary symplectic matrix given by
\begin{equation}
(J\,)^{ij}=\left(
\begin{array}{cc}
0&1\\ -1&0
\end{array}
\right) \ .\label{j}
\end{equation}
Remarkably, Eq.(\ref{poisson1}) does not change under the action
of a symplectic transformation on the phase-space:
\begin{equation}
{\bfm X}\rightarrow{\bfm X^\prime}=T\, {\bfm X} \ ,
\end{equation}
where ${\bfm X}^{\rm t}=(x^1,\,x^2)$ and $T\in$ Sp(1) with
$T^i_{\,\,j}$ belonging to the fundamental representation of $T$.
This property follows from the relation
\begin{equation}
J^{rs}\, T^i_{\,\,r}\, T^j_{\,\,s}=J^{ij} \ .\label{cttc}
\end{equation}
As  is well known Eq. (\ref{poisson1}) can  also be expressed as
\begin{equation}
\Big\{f,\,g\Big\}=\{x^i,\,x^j\}\,\partial_{i}f\,\partial_{j}g
\ ,
\end{equation}
so that, if we know the PB between the generators $x^i$ we
can compute the PB between any pair of functions.

Let us now introduce the $q$-symplectic group Sp$_q$(1), generated
by $\hat T^i_{\,\,j}$, the elements of a $2\times2$ matrix $\hat
T$ belonging to the fundamental representation of
Sp$_q$(1).\\
The $q$-commutativity of the elements of $\hat T$ are controlled by the
$RTT$ relation:
\begin{equation}
R^{ij}_{\,\,\,\,kl}\,\hat T^k_{\,\,r}\,\hat T^l_{\,\,s}= \hat
T^j_{\,\,l}\,\hat T^i_{\,\,k}\,R^{kl}_{\,\,\,\,rs}
 \ ,\label{rtt}
\end{equation}
where the matrix $R$ is defined by
\begin{equation}
(R\,)^{ij}_{\,\,\,\,kl}=\left(
\begin{array}{cccc}
q&0&0&0\\
0&q^{-1}&0&0\\
0&\lambda&q^{-1}&0\\
0&0&0&q
\end{array}
\right) \ ,\label{r}
\end{equation}
with $\lambda=q^{-1}\,(q^2-q^{-2})$ (row and column are numbered
as $11,\,12,\,21,\,22$, respectively). It satisfies the quantum
Yang-Baxter equation
\begin{equation}
R^{ij}_{\,\,\,\,kl}\,R^{kr}_{\,\,\,\,st}\,R^{lt}_{\,\,\,\,uv}=
R^{jr}_{\,\,\,\,lt}\,R^{it}_{\,\,\,\,kv}\,R^{kl}_{\,\,\,\,su} \
,\label{rrr}
\end{equation}
a sufficient condition for the consistency of the RTT relation
(\ref{rtt}). Let us observe that by virtue of the isomorphism
SL$_q$(2)=Sp$_q$(1) the matrix $q^{-1}\,R_q$ coincides with the
matrix $R_{q^2}$ of the GL$_q$(2).\\
The symplectic condition on the element $\hat T^i_{\,\,j}$ is
taken into account by means of a matrix $C$ given by
\begin{equation}
(C)^{ij}=\left(
\begin{array}{cc}
0&q^{-1}\\
-q&0
\end{array}
\right) \ ,\label{c}
\end{equation}
through the relation
\begin{equation}
C^{rs}\,\hat T^i_{\,\,r}\,\hat T^j_{\,\,s}=C^{ij} \ .\label{ctt}
\end{equation}

In analogy with the classical PB, we introduce the
following bilinear operator on the generators $\hat x^i$
\begin{equation}
\Big\{\hat x^i,\,\hat x^j\Big\}_q=(\hat\partial_{r}\hat
x^i)\,\widetilde J^{rs}\,(\hat\partial_{s}\hat x^j) \
,\label{qpoisson}
\end{equation}
where we have defined the unitary $q$-symplectic matrix
$\widetilde J^{rs}$ in
\begin{equation}
(\widetilde J\,)^{ij}=q\,(C)^{ij} \ .\label{qj}
\end{equation}
Eq.(\ref{qj}) is the $q$-analogue of  Eq. (\ref{j}) which
is recovered in the $q\to1$ limit.

By construction, taking into account Eq. (\ref{ctt}),  it
immediately follows that Eq. (\ref{qpoisson}) is invariant under
the action of the $q$-symplectic group Sp$_{q}$(1):
\begin{equation}
\hat x_i\rightarrow \hat x^\prime_i=\hat x_j\,\hat
T^j_{\,\,\,i} \ ,\label{trasf}
\end{equation}
where we assume the commutation between group elements and the
plane elements.

Explicitly Eq. (\ref{qpoisson}) becomes
\begin{equation}
\Big\{\hat x^i,\,\hat x^j\Big\}_q=\hat\partial_x\,\hat
x^i\,\hat\partial_p\,\hat x^j-q^2\,\hat\partial_p\,\hat
x^i\,\hat\partial_x\,\hat x^j \ .\label{qpoisson1}
\end{equation}
It is easy to verify the following fundamental relations
\begin{eqnarray}
&&\Big\{\hat x,\,\hat x\Big\}_q=\Big\{\hat p,\,\hat p\Big\}_q=0 \ ,\\
&&\Big\{\hat x,\,\hat p\Big\}_q=1 \ ,\label{pxp}\\
&&\Big\{\hat p,\,\hat x\Big\}_q=-q^2 \ ,\label{ppx}
\end{eqnarray}
which coincide with the one obtained in Ref.
\cite{Klimek}.\\
In particular, from Eqs. (\ref{pxp}) and (\ref{ppx}) it follows
that Eq. (\ref{qpoisson}) is not antisymmetric. A similar behavior
appears also in the quantum $q$-oscillator theory introduced
earlier.

By means of Eqs.(\ref{x})-(\ref{dp}), a realization of our
generalized $q$-PB can be written as
\begin{equation}
\Big\{f,\,g\Big\}_q={\cal D}_x\,f(x,\,p\,D_x)\,{\cal
D}_p\,g(q\,x,\,p\, D_x)-q^2\,{\cal D}_p\,f(q\,x,\,p\,
D_x)\,{\cal D}_x\,g(x,\,p\, D_x) \ ,\label{qpoisson2}
\end{equation}
where $f$ and $g$ are identified with $x$ or $p$,
respectively.

\section{Conclusions}

In this paper, we have shown that the $q$-calculus plays a crucial
role in the definition of the quantum mechanics of
$q$-oscillators, thermodynamics and in a $q$-classical theory,
defined by means of the introduction of a $q$-PB. We have shown
that the realization of the thermostatistics of $q$-deformed
algebra can be built on the formalism of $q$-calculus and that the
entire structure of thermodynamics is preserved if we use the
(non-commutative) JD instead of the ordinary thermodynamic
derivative. In analogy with quantum group invariance properties of
the quantum $q$-oscillation theory, the $q$-PB has been defined by
assuming the invariance under the action of Sp$_q$(1) group and
its derivatives act on the $q$-deformed non-commutative plane
invariant under GL$_q$(2) transformations. Therefore such a
classical $q$-deformation theory can be seen as the analogue of
$q$-oscillator deformation in the quantum theory. This opens the
possibility of introducing a classical counterpart of the quantum
$q$-deformations and we expect that such a classical $q$-deformed
dynamics can be very relevant in several physical applications, in
a manner similar  to the classical Tsallis' thermostatistics,
based on a deformation of the Boltzmann-Gibbs entropy
\cite{tsallis,tsallis2}.

Although a complete treatment of this matter lies out the scope of
this paper, we would like to state that, upon simple canonical
quantization of the  $q$-classical theory which has been under
investigation, it is possible to obtain, consistently, the
$q$-deformed Heisenberg uncertainty relations and the $q$-deformed
oscillator algebra, introduced in Sec. 2. Finally, for future
investigations we would like to mention some interesting
developments in the $q$-classical harmonic oscillators,
equipartition theorem, Euler's theorem, Liouville theorem and in 
$q$-deformed classical thermodynamics.


\end{document}